\documentclass[11pt]{article} 
\usepackage{hyperref} \pdfoutput=1
\usepackage{graphicx}

\begin{document} 
\title{Secondary vortices in swirling flow}
\author{Radu Cazan and Cyrus K. Aidun \\ \\\vspace{6pt} George W. Woodruff School of Mechanical Engineering, \\ Georgia Institute of Technology, Atlanta, GA 30318, USA}

\maketitle


			\begin{figure}[htbp]
			\centering
				  \includegraphics[width=0.99\textwidth]{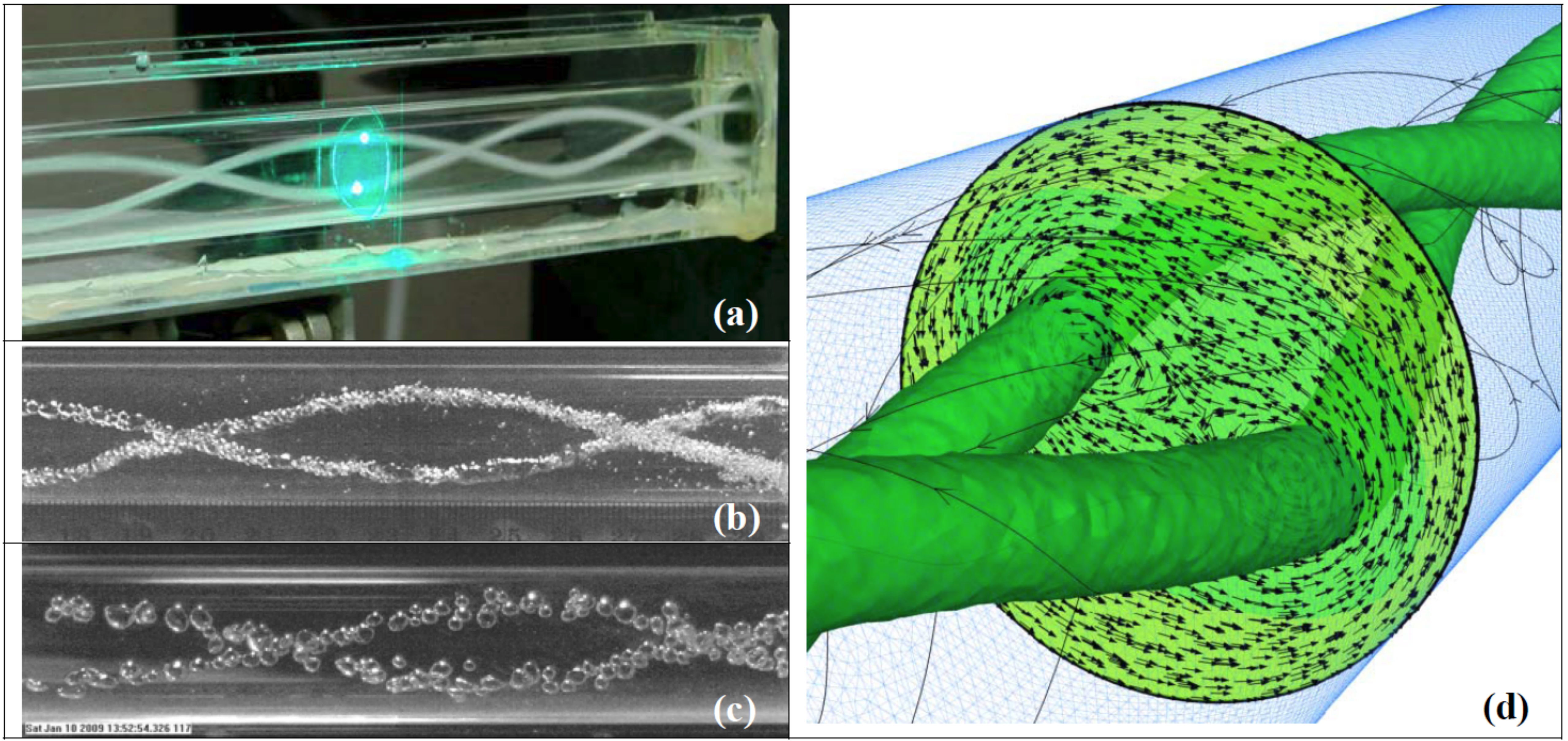}
			\caption{Helical vortices induced by 180$^{o}$ twisted tapes (the flow is from right to left): (a) real time visualization at Re = 7.7$\times$$10^{4}$, (b) high speed camera view at Re = 7.7$\times$$10^{4}$, (c) high speed camera view at Re = 2.5$\times$$10^{4}$, (d) numerical simulation at Re = 7.7$\times$$10^{4}$.}
				\label{fig}
			\end{figure}

Twisted tapes are used to induce swirling flow and improve mixing. The flow induced by a 180$^{o}$ twisted tape with length (pitch) 60 mm and diameter 25.4 mm in a circular pipe was investigated using Laser Doppler Velocimetry (LDV) measurements. Tangential velocity profiles downstream of the twisted tape swirler were measured at multiple locations along the pipe axis, across the horizontal diameter of the pipe.

The profiles showed an unexpected transition along the pipe axis from regular swirling flow to an apparent counter-rotation near the pipe axis, and then reverting back to regular swirling flow1. Injecting fine air bubbles into the flow showed the existence of two co-rotating helical vortices superimposed over the main swirling flow\cite{Caz:09}. The close proximity of the two co-rotating vortices creates the local reversing flow at the pipe centerline.

Figure 1(a) shows a real time view of the flow at Re = 7.7$\times$$10^{4}$. The laser sheet marks the positions of the helical vortices in the cross-section. Figures 1(b) and 1(c) show high speed camera views of the flow at Re = 7.7$\times$$10^{4}$ and Re = 2.5$\times$$10^{4}$, respectively. 

A steady state numerical simulation added more details for the complex flow. Figure 1(d) shows cross-section velocity vectors, low pressure isobar surfaces and streamlines at Re = 7.7$\times$$10^{4}$.

\bibliographystyle{plain}
\bibliography{cazan.bib}

\end{document}